\documentclass[twocolumn,
runinaddress,
preprintnumbers,
noeprint,
aps,amsmath,amssymb,
]{revtex4-2}

\usepackage[utf8]{inputenc}

\usepackage{amsmath}
\usepackage{amssymb}
\usepackage{graphicx}
\usepackage{esint}
\usepackage{hyperref}
\usepackage{multirow}
\usepackage{color}
\usepackage[makeroom]{cancel}
\usepackage[normalem]{ulem}

\begin{document}

\title{Evolution equation for elastic scattering of hadrons}

\author{Hiren Kakkad}
\email{kakkad@agh.edu.pl}
\author{Anderson Kendi Kohara}
\email{kohara@agh.edu.pl}
\author{Piotr Kotko}
\email{pkotko@agh.edu.pl}

\affiliation{ AGH University Of Science and Technology, Faculty of Physics and Applied Computer Science, \\ Mickiewicza 30, 30-059 Krak\'ow, Poland
}


\begin{abstract}

   We turn high energy elastic scattering of hadrons into an initial value problem using an evolution equation based on the Regge Field Theory, which 
   has a form of the complex nonlinear reaction-diffusion equation, with time being played by the logarithm of energy.
   The initial conditions are provided by the data-driven models for the real and imaginary parts of the amplitude. Numerical calculations of pp differential cross sections and forward quantities for LHC energies agree very well with experimental data extending up to, and including the diffractive cone.
   Furthermore, we show that at current accessible energies the non-linear effects play an important role, as the impact parameter space profiles approach the unitarity bound. 
   The equation also predicts some other effects discussed in the literature, like the hollowness of nuclear matter or existence of stationary points in momentum transfer $t$. 
\end{abstract}

\maketitle

\section{Introduction}

High energy relativistic scattering of hadrons can be classified into two major types: elastic and inelastic. 
If a large momentum scale is present,
the latter 
can be treated within perturbative Quantum Chromodynamics (QCD) and  factorization theorems.
The elastic scattering of hadrons, on the other hand, belongs to a non-perturbative domain of QCD, except for  very large momentum transfers, where perturbative techniques may be applied. 
Despite the apparent simplicity of the process, which depends only on two kinematical variables $s$ (the center of mass energy squared) and $t$ (the transferred momentum squared), it is still considered an open problem, as there is no convincing description of elastic scattering in terms of QCD. 

At  $\sqrt{s}\sim$ few GeV the elastic scattering data can be described relatively well by the Regge phenomenology \cite{donnachie_dosch_landshoff_nachtmann_2002}, which represents the scattering amplitude using the partial wave expansion and singularities in the complex {\it l}-plane. 
 The linear relation between
$l$ and $t$, called the Regge trajectory, can be attributed to an exchange of family of meson in $t$ channel. 
However, for large energies (ISR domain and beyond), these regular reggeon linear trajectories fail to describe the rise of the total cross section, as well as the shrinkage of the diffractive cone. 
Thus, one introduces  another effective trajectory with a lower slope and positive intercept. This effective object is the so-called \emph{Pomeron}, a pseudo-particle with isospin zero and even $\mathcal{C}$-parity.

The perturbative QCD predicts the Pomeron as a  color singlet composed of two gluons.
Its energy evolution is given by the Balitsky-Fadin-Kuraev-Lipatov (BFKL) equation \cite{Fadin:1975cb,*Kuraev:1977fs,*Balitsky:1978ic} (see also \cite{BFKL-review:1997} for a review), 
which predicts a power like growth of the gluon density with energy.
However, for asymptotic energies, the power growth has to be tamed, due to the Froissart bound based on the unitarity of the scattering matrix and the finiteness of strong interaction range.
From the perturbative perspective, the gluon density gets so large that fusion of gluons becomes important, leading to the gluon saturation phenomenon \cite{Gribov:1983ivg}. 
Saturation can be incorporated in the non-linear extension of the BFKL --  the 
Balitsky-Kovchegov (BK) equation \cite{Balitsky:1995ub,Kovchegov:1999yj,Marquet:2005qu},
which is a mean field approximation to an infinite set of coupled differential equations for Wilson line correlators known as  Balitsky-Jalilian-Marian-Iancu-McLerran-Weigert-Leonidov-Kovner (B-JIMWLK) equation \cite{Balitsky:1995ub,JalilianMarian:1997jx,*JalilianMarian:1997gr,*JalilianMarian:1997dw,*Kovner:2000pt,*Kovner:1999bj,*Weigert:2000gi,*Iancu:2000hn,*Ferreiro:2001qy}.

Despite the fact that these sophisticated QCD evolution equations  had a tremendous impact on both theoretical and phenomenological studies of high energy processes, in particular processes not involving hadrons (onium-onium scattering \cite{Navelet:1997tx}), they are not appropriate to describe elastic scattering experimental data \cite{Ewerz:2004}.
The perturbative treatment leads to the conformal, impact-parameter-independent evolution kernel (the $b$ appears only in the Pomeron amplitude in combination with the target transverse size -- such $b$ dependent solution was discussed in \cite{Golec-Biernat:2003naj}).
The lack of $b$-space diffusion was recently addressed in \cite{Levin:2020}, but the equations have not been tested phenomenologically.

In this work, we formulate an evolution equation for the {\it complex} elastic scattering amplitude that successfully describes the data and has some resemblance to the aforementioned perturbative equations. Before we introduce and motivate our equation, let us briefly outline the main concepts.
The diffusion in impact parameter space was first discussed by Gribov \cite{Gribov-diffusion}.
In the 60's, he also developed the Reggeon Calculus \cite{Reggeon-Calculus:1967}, where the Pomeron was treated as a quasi-particle exchanged in the $t$ channel, which ultimately lead to the formulation of the Regge Field Theory (RFT). In \cite{abarbanel1975reggeon,Amati:1976,Allesandrini:1976}
a Lagrangian with triple interaction vertex was 
studied to describe the Pomeron interactions.
It is interesting to note that, already in 70', a rapidity evolution equation for the Pomeron correlation functions, preserving diffusion and containing the non-linear quadratic term, was derived in the semi-classical approximation \cite{Alessandrini:1977}. 
This equation has a form of the Fisher-Kolmogorov-Petrovsky-Piscounov (FKPP)   equation (the reaction-diffusion equation) \cite{FKPP:1937_A,*FKPP:1937_B}, which is known to be related to the QCD BK equation \cite{Munier-Peschanski:2003}. 
 At that time there were no experimental data at high energies that could utilize the non-linear phenomena encoded in these equations. Nowadays, in the LHC era, the energies involved are adequate not only to study the non-linear phenomena that elastic amplitudes must contain, but also to study more subtle effects coming from inclusion of the Odderon fields and higher order Pomeron couplings.
 The Odderon is a fundamental prediction of perturbative QCD, which is a color singlet state of gluons with odd parity. 
Recently, the D0 and Totem collaborations, studying the difference between $\mathrm{pp}$ and  $\mathrm{p\bar{p}}$ cross sections at different energies, concluded that an odd amplitude is necessary to explain the data leading to the possible Odderon discovery at the LHC. \cite{Odderon:2021}.

In the present work, based on \cite{Alessandrini:1977}, we derive a nonlinear evolution equation for the complex elastic amplitude and confront it with up-to-date experimental data. The evolution equation is somewhat similar the one presented in Ref.~\cite{Peschanski:2009}, where analytic properties of the solutions to the real FKPP equation with a noise term were investigated, independently of the initial conditions. In our work, on the other hand, we solve the full complex equation, tying together the real and imaginary amplitudes, using the data-driven phenomenological models \cite{kohara:2014, BSW:1979} as initial conditions.
We show that our complex equation describes the data exceptionally well up to the dip-bump region in the differential elastic cross section as well as  the quantities in the forward regime.


\section{Evolution equation for elastic  amplitude}
\vspace{5pt}
The relation between the S-matrix and the elastic scattering amplitude reads
\begin{equation}
    \mathcal{S}(s,t) = 1+i\mathcal{T}(s,t) \,,
\end{equation}
 where the transition amplitude $\mathcal{T}$ has real and imaginary parts
\begin{equation}
    \mathcal{T}=T_R+iT_I \,.
\end{equation}
We define the impact parameter space amplitude as the Fourier transform
of the transverse components of the momentum transfer $q$, $t\simeq-\mathbf{q}^2$,
\begin{equation}
    \widetilde{\mathcal{T}}(\tau,\mathbf{b})=
    \int\!d^2\mathbf{q}\, e^{-i\mathbf{b}\cdot\mathbf{q}}\, \mathcal{T}(\tau,-{\bf q}^2) \,,
    \label{eq:b-space_amp}
\end{equation}
where $\tau \simeq\log{s}$ is the rapidity and the bold symbols represent transverse vectors.

In the Regge theory the energy growth of the amplitude is attributed to the Pomeron exchange:
\begin{equation}
    \mathcal{T}(s,t) = g_a(t)\,g_b(t)\,e^{[\alpha_P(t)-1]\log s}\,\Big(i-\cot\frac{\pi\alpha_P(t)}{2}\Big) \,,
    \label{eq:pomeron-amplitude}
\end{equation}
where $\alpha_P(t)=1+\epsilon_0+\alpha^\prime\,t~$ is the standard linear Pomeron trajectory with the intercept $\epsilon_0$ and the slope $\alpha'$. The functions $g_i$ are the hadronic form factors.

It is  easy to check that the $b$-space amplitude, in the large $b$ limit, satisfies the following diffusion equation 
\begin{equation}
    \frac{\partial\widetilde{\mathcal{S}}(\tau,\mathbf{b})}{\partial \tau}
    = \left(\alpha' \nabla_{\mathbf{b}}^2 - \epsilon_0\right) \widetilde{\mathcal{S}}(\tau,\mathbf{b})\,~.
    \label{eq:diffiusion}
\end{equation}

 Eq.~\eqref{eq:diffiusion}, however,  violates the Froissart bound and, furthermore,  its solutions are in contradiction with high energy LHC data.
To correct that, we use the RFT approach with the following Lagrangian density \cite{abarbanel1975reggeon}
\begin{equation}
\mathcal{L} = \frac{1}{2}p\overleftrightarrow{\partial}_{\tau} q+\alpha'\nabla_b q\cdot\nabla_b p -\epsilon_0\,p\,q+\lambda\,p\,q\,(p+q) \,,
    \label{Lagrangian:RFT}
\end{equation}
where the fields $q$ and $p$ are related to Gribov's  Pomeron fields, $q=i\overline{\Psi}$, $p=i\Psi$ and depend both on $b$ and  $\tau$. Above, $\lambda$ is the triple Pomeron coupling.
 As pointed out in \cite{Amati:1976}, in addition to the zero energy ground state $|\phi_0\rangle$, the Hamiltonian acquires a non zero energy state $|\phi_1\rangle$, which for large $\tau$ and positive intercept $\epsilon_0$  approaches a coherent state. A general $\tau$ dependent state is constructed as $|\psi(\tau)\rangle=\exp(-\hat{A}(\tau))|\phi_0\rangle$, where  the operator $A(\tau)$ is given by an infinite sum of $(n+1)$-point correlation functions convoluted with the Pomeron creation operators $\hat{q}$:
\begin{equation}
\hat{A}(\tau)=\sum_{n=1}^ {\infty}\frac{1}{n!}\int\!\! d^2\mathbf{b}_1...d^2\mathbf{b}_n\,\hat{q}(\mathbf{b}_1)...\hat{q}(\mathbf{b}_n)\,G_n(\tau,\mathbf{b}_1,..\mathbf{b}_n)~.
    \label{A_operator}
\end{equation}
Eq.~(\ref{A_operator}) inserted to the Schrödinger equation
$
   \partial_{\tau}|\psi(\tau)\rangle =- H\,|\psi(\tau)\rangle
    \label{Schrodinger:eq}
$,
gives an infinite set of coupled differential equations for the  correlators $G_n$. In the semi-classical approximation it reduces to an uncoupled set of equations. In particular, for the two point correlation function one gets \cite{Allesandrini:1976}
\begin{equation}
    \frac{\partial G_1(\tau, \mathbf{b})}{\partial \tau}
    = \left(\alpha' \nabla_{\mathbf{b}}^2 +\epsilon_0\right) G_1(\tau, \mathbf{b})-\lambda G_1(\tau, \mathbf{b})^2\,.
    \label{eq:FKPP_correlator}
\end{equation}
Since the energy dependence of the elastic amplitude is driven by the one-Pomeron exchange, cf. Eq.~\eqref{eq:pomeron-amplitude}, the evolution equation for the two-point correlator can be readily transformed into the equation for the  amplitude itself and thus the S-matrix. Because hadrons act as sources in RFT, the $b$-dependent hadron state is simply represented by a coherent state $|p(\mathbf{B})\rangle=\exp(-\int\!d^2b\,f(\mathbf{b}-\mathbf{B})\,\hat{q}(\mathbf{b}))|\phi_0\rangle$ with the profile function $f$ peaked around zero. The source function $f$, together with $G_1(0,\mathbf{b})$, provide the initial conditions and can be complex valued. Putting these ingredients together, we obtain the following equation for the S-matrix
\begin{equation}
    \frac{\partial\widetilde{\mathcal{S}}(\tau,\mathbf{b})}{\partial \tau}
    = \left(\alpha' \nabla_{\mathbf{b}}^2 - \epsilon_0\right) \widetilde{\mathcal{S}}(\tau,\mathbf{b})+\lambda\widetilde{\mathcal{S}}(\tau,\mathbf{b})^2\,.
    \label{eq:FKPP}
\end{equation}
This equation resembles the FKPP equation \cite{FKPP:1937_A,*FKPP:1937_B}. However, since  $\widetilde{\mathcal{S}}$ is a complex quantity  the above equation provides a set of two real coupled differential equations for the real and imaginary amplitudes.
An interesting fact is that these two equations 
 \begin{eqnarray}
\label{evolution-Pomeron}
\frac{\partial \widetilde{T_I}}{\partial \tau} = \alpha^\prime\frac{\partial^2 \widetilde{T_I}}{\partial b^2}+\epsilon_0\left[\widetilde{T_I}(1-\lambda\widetilde{T_I})+\lambda\widetilde{T_R}^2\right]~ ,
\end{eqnarray}
and
\begin{eqnarray}
\label{evolution-Odderon}
\frac{\partial \widetilde{T_R}}{\partial \tau} = \alpha^\prime\frac{\partial^2 \widetilde{T_R}}{\partial b^2}+\epsilon_0\,\widetilde{T_R}\, (1-2\lambda\,\widetilde{T_I})~ 
\end{eqnarray}
have a very similar structure to the Pomeron and Odderon energy evolution equations of the BK type \cite{Munier-Peschanski:2003,Odderon:2003}.

In order to use the above equations one needs  appropriate initial conditions, which specify the impact parameter profile functions for $\widetilde{T_I}$ and $\widetilde{T_R}$, at a given initial energy. Once these are known, for instance, from phenomenological models/parametrizations, they provide both the energy dependence and impact parameter evolution of the amplitudes at any energy.

Eq.~\eqref{eq:FKPP} is derived solely from the triple-Pomeron interactions within the RFT, and as discussed before, does not include the Odderon fields and higher order Pomeron couplings in the evolution (it may include them however, indirectly, in the initial condition, as both real and imaginary parts have to be provided).
Nevertheless, Eq.~\eqref{eq:FKPP} captures the key elements of the experimental data when supplied with appropriate initial profiles, leaving space for inclusion of the Odderon fields and other higher order non-linear  terms.

\section{Numerical Results}
\vspace{5pt}
In this section we solve the nonlinear equation~\eqref{eq:FKPP}, equivalent to the set of equations \eqref{evolution-Pomeron}-\eqref{evolution-Odderon}, numerically and confront it with the existing $\mathrm{pp}$  elastic scattering  data. Moreover, we provide predictions for energies not accessible today, but likely to be explored by the future particle colliders or cosmic ray experiments.

In our normalization the differential cross section reads:
\begin{equation}
\frac{d\sigma}{dt} = \pi\,(\hbar c)^2\Big[T_I(s,t)^ 2+T_R(s,t)^ 2\Big]~,
\label{diff-cross}
\end{equation}
where $(\hbar c)^2 =  0.389379 \,\, \mathrm{mb\, GeV^2}$.
For the sake of simplicity, in our analysis we disregard the Coulomb interaction and consequently its interference with the nuclear amplitude. 
The forward quantities, obtained in the limit of $t=0$, include the total cross section $\sigma_{\rm tot}$ via the optical theorem, the ratio $\rho$ of the real and imaginary parts, and the imaginary and real slopes $B_I$, $B_R$. These are defined respectively as follows:
\begin{equation}
\sigma_{\rm tot} =4\,\pi\,(\hbar c)^2\,T_I(s,0) ~,
    \label{total-sigma}
\end{equation}
\begin{equation}
\rho = \frac{T_R(s,0)}{T_I(s,0)}~,
    \label{rho}
\end{equation}
\begin{equation}
B_{I,R} = \frac{2}{T_{I,R}(s,t)}\frac{d T_{I,R} (s,t)}{dt}\Bigg|_{t=0}~.
    \label{slopes}
\end{equation}

The initial conditions for the differential equation have to be set up by a measurement at lower energies  through extracted $b$-space profiles, both for the real and imaginary parts. In the present work we use the Kohara-Ferreira-Kodama (KFK) model~\cite{kohara:2014,Kohara:2021}, and independently the Bourrely-Soffer-Wu (BSW) model \cite{BSW:1979} to test the initial condition dependence of the predictions. The main advantage of these models is that they possess analytic forms in $b$-space. 

It is important to stress that the initial conditions require $b$-space profiles at certain fixed initial energy.
Our phenomenological studies show that our evolution equation
is suitable for large energies (beyond ISR). 
We observe that the  real part of the evolution equation is  monotonically reduced with the increasing energy, which, consequently, causes a reduction in the $\rho$ parameter.
 Therefore, we choose to start with the initial condition at $\sqrt{s}=$ 500 GeV, since at this energy, according to the dispersion relations expectations for a given total cross section parametrization (see PDG \cite{PDG:2020}), the $\rho$ value should be close to its maximum. 

In Table~\ref{tab:parameters} we show the parameters of the evolution equation, $\alpha'$, $\epsilon_0$ and $\lambda$, determined in such a way that our solution-based differential cross-section -- for some higher energy, in our case 7 TeV -- fits the experimental data. 
 These values are rather stable for different initial conditions and used for all other energies.  It is remarkable that the Pomeron intercept $\epsilon_0$ is close to the standard nonperturbative intercept $\epsilon_{\mathcal{P}}=0.096$. The Pomeron slope (the diffusion coefficient) is close to  0.1 GeV$^{-2}$,  which is   below the standard value 0.25 GeV$^{-2}$, but the possibility of this reduction was predicted in the 70's \cite{Amati:1975}. 
We also observe the stability of the triple Pomeron coupling.

\begin{table}[h]
\begin{tabular}{cccc}
      \qquad\qquad\qquad               & $\alpha' \,\, ( \mathrm{GeV^{-2}})$                        & $\epsilon_0$& $\lambda$   \\ \hline\hline
KFK & 0.105 & 0.129 & 0.712 \\ 
BSW & 0.090 & 0.140 & 0.820 \\ \hline
\end{tabular}
\caption{The parameters $\alpha'$, $\epsilon_0$ and $\lambda$ are presented separately for the KFK and the BSW initial conditions.
}
\label{tab:parameters}
\end{table}

\begin{figure}
    \centering
 \includegraphics[width=8.5cm]{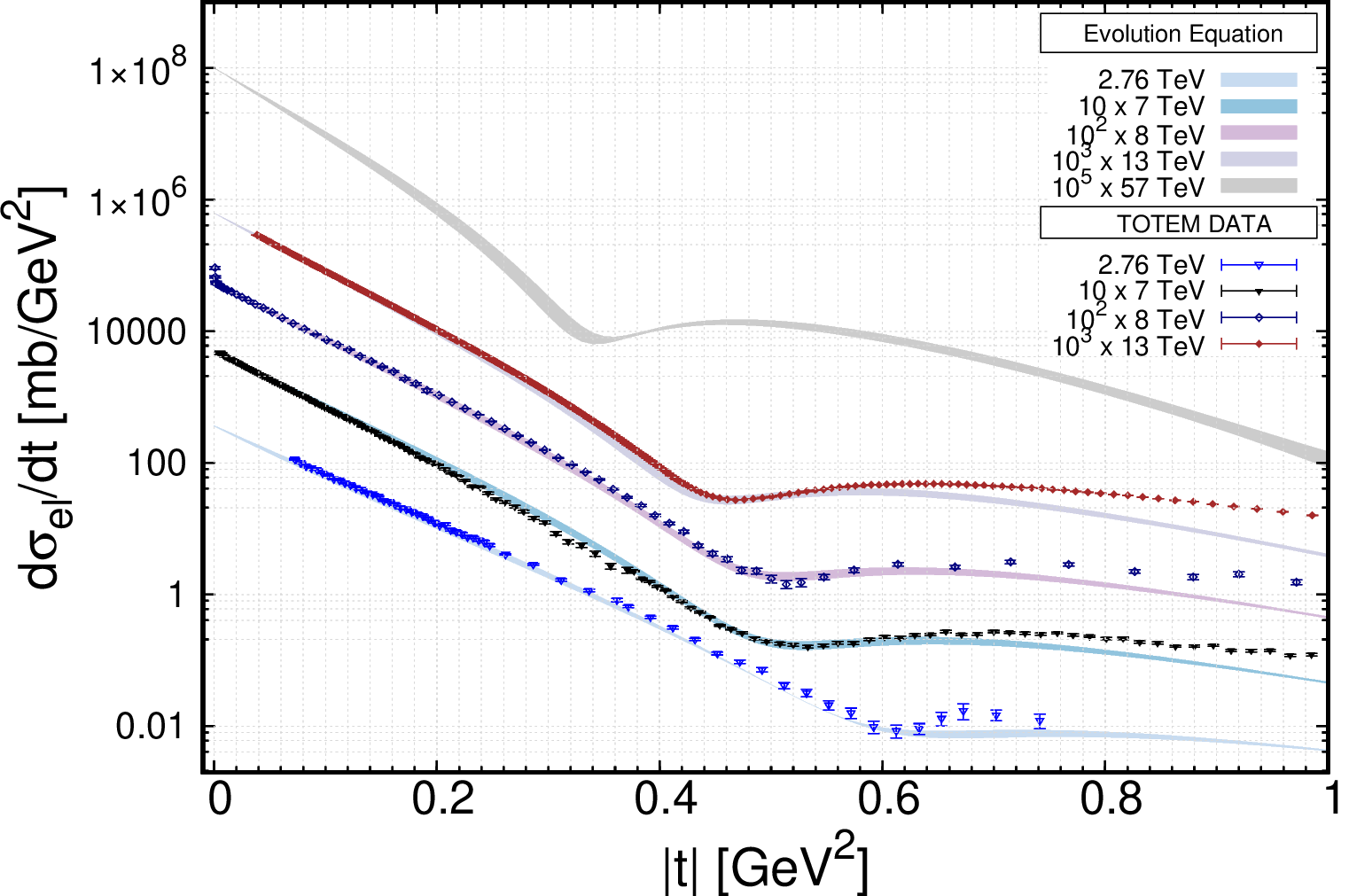}
    \caption{\small The points represent the TOTEM proton-proton elastic differential cross-section measured at 2.76. 7, 8 and 13 $\mathrm{TeV}$ with statistical  uncertainties (bars) in the $|t|$ range $0 < |t| < 1$ $\mathrm{GeV}^2$. The colored bands represent the predictions made using the coupled  non-linear evolution equation Eq~(\ref{evolution-Pomeron})-(\ref{evolution-Odderon})  for the KFK and BSW $b$-space initial profiles. }
    \label{fig:diff_cross}
\end{figure}

For the available LHC energies and low to moderate $t$  we observe a good match between our predictions and the  TOTEM  data for the differential cross section. The bands shown in Fig.~\ref{fig:diff_cross} result from the difference in the predictions for the KFK and BSW initial conditions. For a given value of $t$, the lower limit of the band is defined by the minimum of the predictions for the two initial conditions, while the upper limit is defined by the maximum of the two. 
For higher values of $t$, including the dip and the bump regions, the positions and magnitudes also match  well with the experimental data. 
 Quantitatively, the calculated dip positions shown in Table~\ref{tab:table_In} are in well agreement with TOTEM measurements. 
Beyond the bump, our solutions go below the experimental data.
Since this region is still far from the perturbative tail in terms of QCD,
in our opinion this could be explained with the inclusion of other interaction terms in the Lagrangian of RFT, which will be investigated in future work.

\begin{figure}[h]
    \centering
 \includegraphics[width=8.5cm]{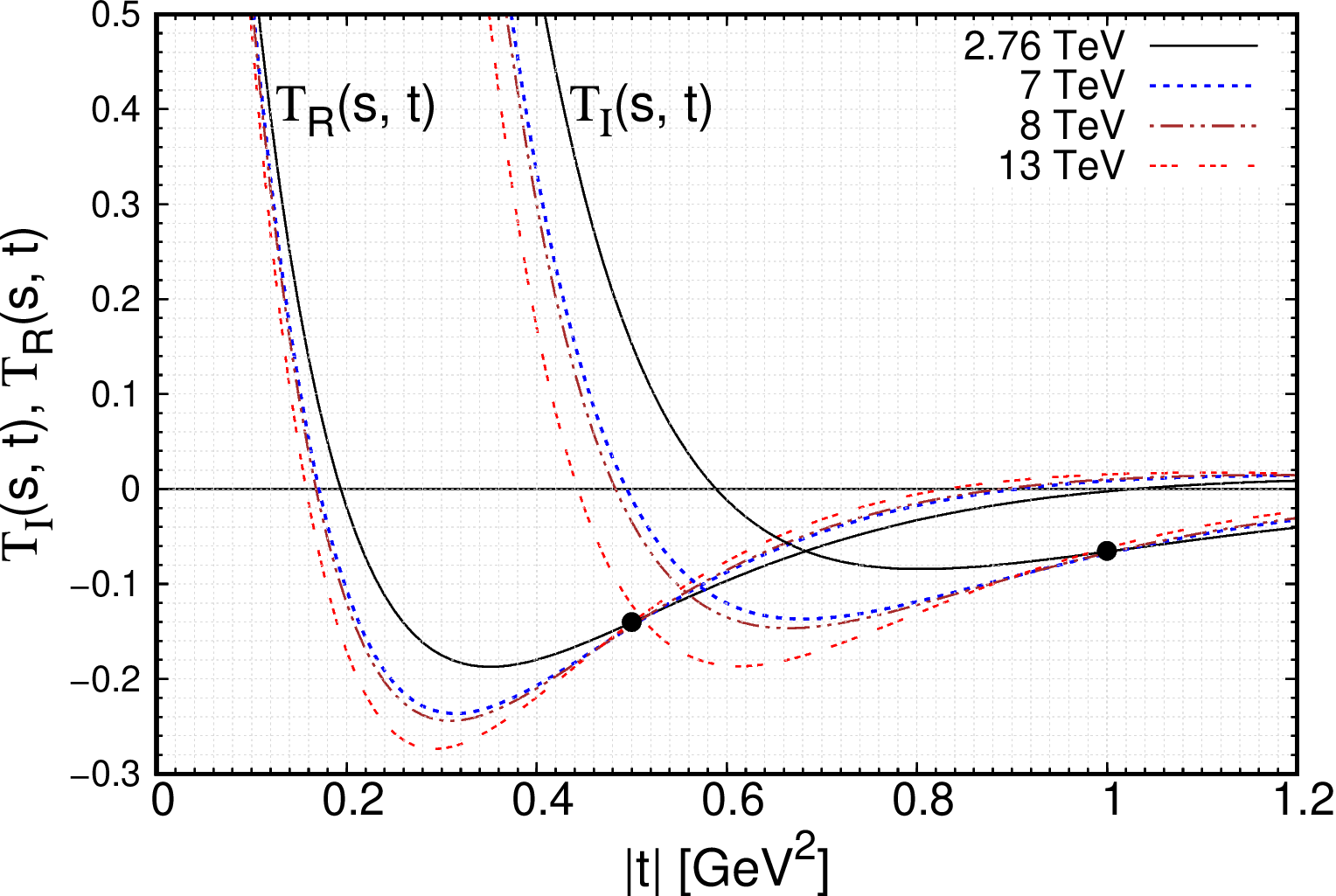}
    \caption{\small 
     Real and imaginary pp amplitudes  obtained numerically for the KFK initial conditions for the LHC energies. }
    \label{fig:amplitudes_t}
\end{figure}

The real and imaginary amplitudes in the $t$-space exhibit similar analytic behaviour compared to other models available in the literature \cite{Kohara:2021, BSW:1979, Selyugin:2015, Csorgo:2021}.
As can be seen in Fig.~\ref{fig:amplitudes_t}, the real part has the first zero before the zero of the imaginary part. This zero approaches the origin  before the zero of the imaginary part with the increasing energy in accordance with the theorem of Martin \cite{Martin:1997}. On the other hand, the zero of the imaginary part is close to the dip position. The dip structure, i.e. its magnitude and shape, are determined by the interplay of the imaginary and the real part, which is non-zero in this region 
for several energies. 

Another intriguing aspect is the presence of fixed  points (stationary points) for various energies in both real and imaginary amplitudes,
 i.e. the points for which the curves intersect for different energies. 
The first fixed point in the real part is at $|t| =  0.07$ GeV$^2$, and the second is at $|t| =  0.5$ GeV$^2$. For imaginary part, the first fixed point is at $|t| = 0.2 $ GeV$^2$, and the second at $|t| = 1.0 $ GeV$^2$. These fixed points constrain the shape of real and imaginary amplitudes and could indicate the uniqueness of the scattering amplitudes.  Although not exactly at the same positions, these fixed points were also observed by  Csörg{\H{o}} et al. in Ref \cite{Csorgo:2021,Csorgo:2022}. 

In Table~\ref{tab:table_In} we summarize the standard forward quantities  calculated at different energies.
For the total cross section $\sigma_{\rm tot}$, the obtained values for both  initial conditions  agree  with  TOTEM \cite{Totem:total} for different energies, and with AUGER \cite{Auger:2012}, whereas for the ATLAS \cite{Atlas:2014,Atlas:2016} data our result seems to be slightly above. For $\rho$, our results are again substantially within the error limits for the extracted values at  7 and 8~$\mathrm{TeV}$ from TOTEM. For 13~$\mathrm{TeV}$, the prediction for KFK initial condition is within the error limits, whereas for BSW  the value falls slightly outside the error bar. 
We believe that the inclusion of the Coulomb interaction might introduce some corrections due to its interplay with the real nuclear part. Interestingly, the measured slope $B$, which is weighted average between the real $B_R$ and imaginary $B_I$ slopes, is in between the calculated values for both initial conditions. 

\begin{figure}
    \centering
 \includegraphics[width=8.5cm]{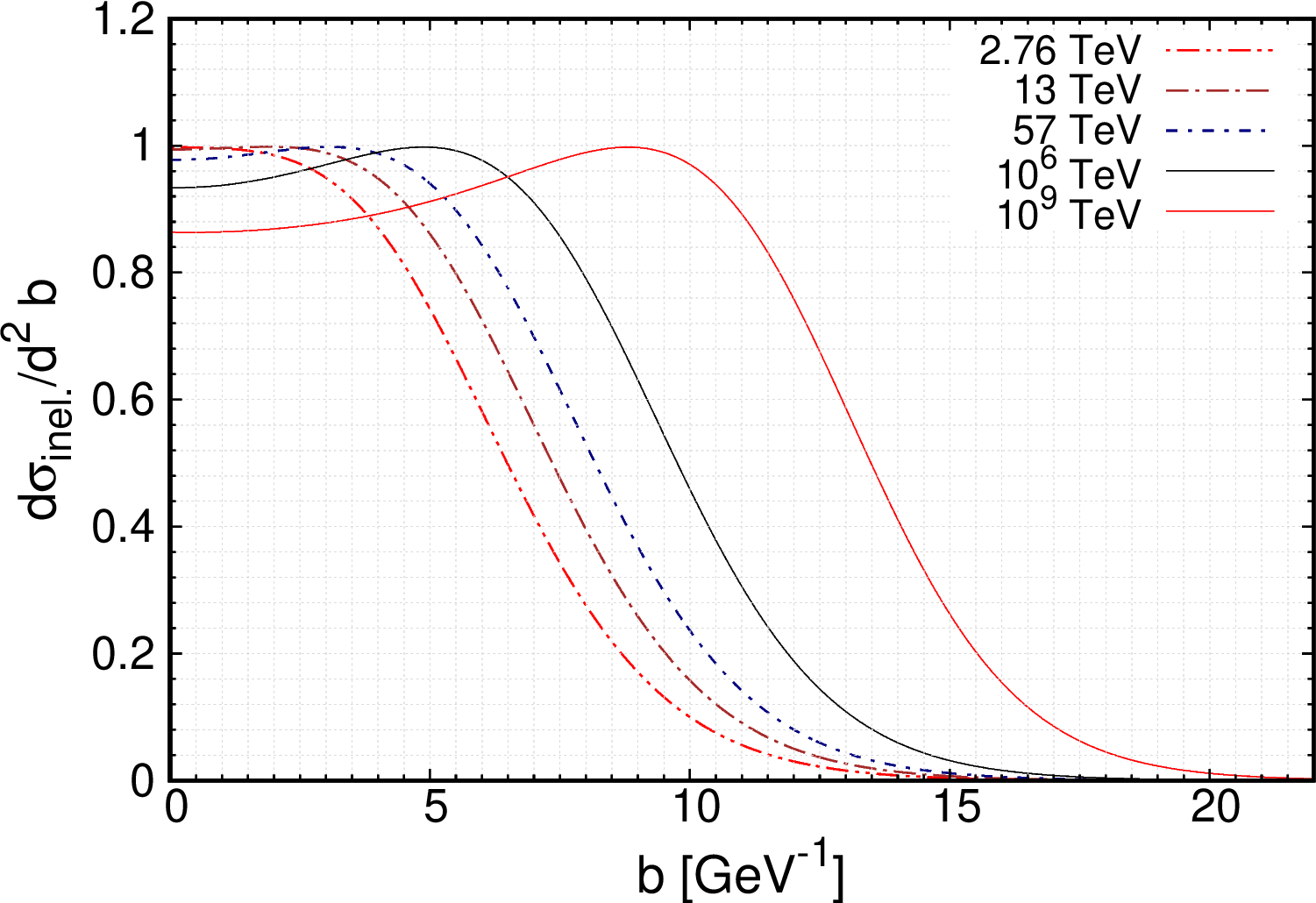}
    \caption{\small 
    Inelastic profiles obtained numerically using the evolution equation Eq.~\eqref{eq:FKPP} for energies between 2.76~TeV  up to  $10^9$~TeV  
    for the KFK initial conditions. 
    }
    \label{fig:Pro_IN}
\end{figure}

Although not observable, the $b$-space profiles are useful to analyze the interaction mechanism. At low energies the diffusive and linear parts of the evolution equation drive the dynamics of the scattering amplitudes. However, at large energies the non-linear effects become important. Due to the presence of the triple Pomeron vertex, the diffusive evolution equation follows a reactive-diffusive system that  contains essentially  two regions: one for stable solution at large $b$ and another for unstable solution at small $b$.  The LHC energies are especially interesting because the transition from the linear regime to the non-linear can be probed. In this energy range the profiles grow with the energy and eventually reach the unitarity bound (similar to the  saturation mechanism for gluons). Once the profiles reach the unitarity bound the only way to allow for the growth of the integrated cross sections is by diffusion for large $b$. For very large energies the profiles behave as traveling waves and we obtain a geometric scaling \cite{Deus:1974}; in other words the scattering process becomes function of a single scaled variable $h(s,t)$.
Similar property was recently discussed in \cite{ CsorgoScaling:2019ewn,Baldenegro:2022xrj}.
Another important observation on the inelastic profile is the existence of a toroidal-like structure \cite{Dremin:2015},  later dubbed as 'hollowness' effect \cite{Broniowiski:2016}, which was also studied in subsequent works \cite{Csorgo:2020,Kohara:2021,Prochzka:2020}. In Fig.~\ref{fig:Pro_IN} we observe this effect starting from 13 TeV and it seems to persist for larger energies. 

\begin{table}[]
\begin{tabular}{ccccc}
                                          & $\sqrt{s}\, \mathrm{[TeV]}$ & $\sigma_{\rm tot}\, \mathrm{[mb]}$ & $\rho$   & $B\, \mathrm{[GeV^{-2}]}$           \\ \hline \hline
\multicolumn{1}{c}{\multirow{5}{*}{\parbox{1.4cm}{KFK \\initial condition}}} & 2.76     & 84.31         & 0.123         & 17.28           \\
\multicolumn{1}{c}{}                     & 7        & 99.07         & 0.117         & 18.47          \\
\multicolumn{1}{c}{}                     & 8        & 101.32        & 0.116         & 18.65        \\
\multicolumn{1}{c}{}                     & 13       & 109.78        & 0.113         & 19.32           \\
\multicolumn{1}{c}{}                     & 57       & 138.32        & 0.105         & 21.55        \\ \hline 
\multicolumn{1}{c}{\multirow{5}{*}{\parbox{1.4cm}{BSW \\initial condition}}} & 2.76     & 83.14         & 0.143         & 19.69          \\
\multicolumn{1}{c}{}                     & 7        & 98.40         & 0.134         & 21.12          \\
\multicolumn{1}{c}{}                     & 8        & 100.73        & 0.132         & 21.34          \\
\multicolumn{1}{c}{}                     & 13       & 109.52        & 0.127         & 22.15         \\
\multicolumn{1}{c}{}                     & 57       & 142.30        & 0.115         & 24.87          \\ \hline
\multicolumn{1}{c}{\multirow{4}{*}{TOTEM}}                                                          & 2.76     &  84.7 ± 3.3    &       -        & 17.1 ± 0.30  \\
\multicolumn{1}{c}{}                                                                              & 7        & 98.0 ± 2.5    & 0.145 ± 0.091 & 19.73 ± 0.40     \\
\multicolumn{1}{c}{}                                                                               & 8        & 101.7 ± 2.9   & 0.12 ± 0.03  & 19.74 ± 0.28 \\
\multicolumn{1}{c}{}                                                                              & 13       & 110.6 ± 3.4   & 0.10 ± 0.01   & 20.40 ± 0.01  \\ \hline
\multicolumn{1}{c}{\multirow{2}{*}{ATLAS}}      & 7    &  95.35 ± 0.38    &       0.14 (fix)        & 19.73 ± 0.14 \\
\multicolumn{1}{c}{}                       & 8        & 96.07 ± 0.18   & 0.136 (fix)  & 19.74 ± 0.05
  \\ \hline
AUGER                          & 57     &  133 ± 29    &     -          &  -  \\\hline
\end{tabular}
\caption{
Forward quantities $\sigma_{\rm tot}$, $\rho$, $B$ calculated using both KFK and BSW initial conditions, for various energies. We also show the available experimental data at LHC energies - TOTEM 2.76 TeV to 13 TeV \cite{Totem:total,Antchev_2020,*TOTEM:7TeV,*TOTEM:8TeV,*TOTEM:13TeV,*Totem:8T} (and the references therein), ATLAS 7 and 8 TeV \cite{Atlas:2014,Atlas:2016} -  and Cosmic Ray experiment - AUGER \cite{Auger:2012}. The $\rho$ values for ATLAS are fixed according to COMPETE parametrization \cite{PDG:2020}. }
\label{tab:table_In}
\end{table}

\section{Summary}
\label{sec:Summary}

Elastic scattering of hadrons is one of the most basic processes, yet it is incredibly rich in complicated dynamics, which is still not completely understood. In the present work we change radically the way one can approach this fundamental problem applying an evolution equation for complex amplitudes. The equation has its origin in dynamics of Pomeron interactions and has the potential to include the Odderon field plus higher order interaction terms. Using realistic models as initial conditions we describe the Large Hadron Colider data in a broad kinematic range and we are able to extrapolate our predictions to cosmic ray energies. 

\section{Acknowledgments}
We thank T.  Csörg{\H{o}},  I. Szanyi, L. Jenkovszky, W. Broniowski, J. Procházka, D. Kroff and A. Stasto for discussions. We thank E. Ferreira for reading the manuscript and for discussions. 
H.K. is supported by the National Science Centre, Poland grant no. 2021/41/N/ST2/02956.  A.K.K. is supported  by the National Science Centre in Poland, grant no. 2020/37/K/ST2/02665 and P.K. is supported by the National Science Centre, Poland grant no. 2018/31/D/ST2/02731. The research leading to these results has received funding from the Norwegian Financial Mechanism 2014-2021.

\bibliographystyle{unsrt} 

\bibliographystyle{apsrev4-2.bst}
\bibliography{references}

\begin{thebibliography}{58}%
\makeatletter
\providecommand \@ifxundefined [1]{%
 \@ifx{#1\undefined}
}%
\providecommand \@ifnum [1]{%
 \ifnum #1\expandafter \@firstoftwo
 \else \expandafter \@secondoftwo
 \fi
}%
\providecommand \@ifx [1]{%
 \ifx #1\expandafter \@firstoftwo
 \else \expandafter \@secondoftwo
 \fi
}%
\providecommand \natexlab [1]{#1}%
\providecommand \enquote  [1]{``#1''}%
\providecommand \bibnamefont  [1]{#1}%
\providecommand \bibfnamefont [1]{#1}%
\providecommand \citenamefont [1]{#1}%
\providecommand \href@noop [0]{\@secondoftwo}%
\providecommand \href [0]{\begingroup \@sanitize@url \@href}%
\providecommand \@href[1]{\@@startlink{#1}\@@href}%
\providecommand \@@href[1]{\endgroup#1\@@endlink}%
\providecommand \@sanitize@url [0]{\catcode `\\12\catcode `\$12\catcode
  `\&12\catcode `\#12\catcode `\^12\catcode `\_12\catcode `\%12\relax}%
\providecommand \@@startlink[1]{}%
\providecommand \@@endlink[0]{}%
\providecommand \url  [0]{\begingroup\@sanitize@url \@url }%
\providecommand \@url [1]{\endgroup\@href {#1}{\urlprefix }}%
\providecommand \urlprefix  [0]{URL }%
\providecommand \Eprint [0]{\href }%
\providecommand \doibase [0]{https://doi.org/}%
\providecommand \selectlanguage [0]{\@gobble}%
\providecommand \bibinfo  [0]{\@secondoftwo}%
\providecommand \bibfield  [0]{\@secondoftwo}%
\providecommand \translation [1]{[#1]}%
\providecommand \BibitemOpen [0]{}%
\providecommand \bibitemStop [0]{}%
\providecommand \bibitemNoStop [0]{.\EOS\space}%
\providecommand \EOS [0]{\spacefactor3000\relax}%
\providecommand \BibitemShut  [1]{\csname bibitem#1\endcsname}%
\let\auto@bib@innerbib\@empty
\bibitem [{\citenamefont {Donnachie}\ \emph {et~al.}(2002)\citenamefont
  {Donnachie}, \citenamefont {Dosch}, \citenamefont {Landshoff},\ and\
  \citenamefont {Nachtmann}}]{donnachie_dosch_landshoff_nachtmann_2002}%
  \BibitemOpen
  \bibfield  {author} {\bibinfo {author} {\bibfnamefont {S.}~\bibnamefont
  {Donnachie}}, \bibinfo {author} {\bibfnamefont {G.}~\bibnamefont {Dosch}},
  \bibinfo {author} {\bibfnamefont {P.}~\bibnamefont {Landshoff}},\ and\
  \bibinfo {author} {\bibfnamefont {O.}~\bibnamefont {Nachtmann}},\ }\href
  {https://doi.org/10.1017/CBO9780511534935} {\emph {\bibinfo {title} {Pomeron
  Physics and QCD}}},\ Cambridge Monographs on Particle Physics, Nuclear
  Physics and Cosmology\ (\bibinfo  {publisher} {Cambridge University Press},\
  \bibinfo {year} {2002})\BibitemShut {NoStop}%
\bibitem [{\citenamefont {Fadin}\ \emph {et~al.}(1975)\citenamefont {Fadin},
  \citenamefont {Kuraev},\ and\ \citenamefont {Lipatov}}]{Fadin:1975cb}%
  \BibitemOpen
  \bibfield  {author} {\bibinfo {author} {\bibfnamefont {V.~S.}\ \bibnamefont
  {Fadin}}, \bibinfo {author} {\bibfnamefont {E.}~\bibnamefont {Kuraev}},\ and\
  \bibinfo {author} {\bibfnamefont {L.}~\bibnamefont {Lipatov}},\ }\href
  {https://doi.org/10.1016/0370-2693(75)90524-9} {\bibfield  {journal}
  {\bibinfo  {journal} {Phys. Lett. B}\ }\textbf {\bibinfo {volume} {60}},\
  \bibinfo {pages} {50} (\bibinfo {year} {1975})}\BibitemShut {NoStop}%
\bibitem [{\citenamefont {Kuraev}\ \emph {et~al.}(1977)\citenamefont {Kuraev},
  \citenamefont {Lipatov},\ and\ \citenamefont {Fadin}}]{Kuraev:1977fs}%
  \BibitemOpen
  \bibfield  {author} {\bibinfo {author} {\bibfnamefont {E.~A.}\ \bibnamefont
  {Kuraev}}, \bibinfo {author} {\bibfnamefont {L.~N.}\ \bibnamefont
  {Lipatov}},\ and\ \bibinfo {author} {\bibfnamefont {V.~S.}\ \bibnamefont
  {Fadin}},\ }\href@noop {} {\bibfield  {journal} {\bibinfo  {journal} {Sov.
  Phys. JETP}\ }\textbf {\bibinfo {volume} {45}},\ \bibinfo {pages} {199}
  (\bibinfo {year} {1977})},\ \bibinfo {note} {[Zh. Eksp. Teor.
  Fiz.72,377(1977)]}\BibitemShut {NoStop}%
\bibitem [{\citenamefont {Balitsky}\ and\ \citenamefont
  {Lipatov}(1978)}]{Balitsky:1978ic}%
  \BibitemOpen
  \bibfield  {author} {\bibinfo {author} {\bibfnamefont {I.~I.}\ \bibnamefont
  {Balitsky}}\ and\ \bibinfo {author} {\bibfnamefont {L.~N.}\ \bibnamefont
  {Lipatov}},\ }\href@noop {} {\bibfield  {journal} {\bibinfo  {journal} {Sov.
  J. Nucl. Phys.}\ }\textbf {\bibinfo {volume} {28}},\ \bibinfo {pages} {822}
  (\bibinfo {year} {1978})},\ \bibinfo {note} {[Yad.
  Fiz.28,1597(1978)]}\BibitemShut {NoStop}%
\bibitem [{\citenamefont {Forshaw}\ and\ \citenamefont
  {Ross}(2011)}]{BFKL-review:1997}%
  \BibitemOpen
  \bibfield  {author} {\bibinfo {author} {\bibfnamefont {J.~R.}\ \bibnamefont
  {Forshaw}}\ and\ \bibinfo {author} {\bibfnamefont {D.~A.}\ \bibnamefont
  {Ross}},\ }\href@noop {} {\emph {\bibinfo {title} {{Quantum chromodynamics
  and the pomeron}}}},\ Vol.~\bibinfo {volume} {9}\ (\bibinfo  {publisher}
  {Cambridge University Press},\ \bibinfo {year} {2011})\BibitemShut {NoStop}%
\bibitem [{\citenamefont {Gribov}\ \emph {et~al.}(1983)\citenamefont {Gribov},
  \citenamefont {Levin},\ and\ \citenamefont {Ryskin}}]{Gribov:1983ivg}%
  \BibitemOpen
  \bibfield  {author} {\bibinfo {author} {\bibfnamefont {L.~V.}\ \bibnamefont
  {Gribov}}, \bibinfo {author} {\bibfnamefont {E.~M.}\ \bibnamefont {Levin}},\
  and\ \bibinfo {author} {\bibfnamefont {M.~G.}\ \bibnamefont {Ryskin}},\
  }\href {https://doi.org/10.1016/0370-1573(83)90022-4} {\bibfield  {journal}
  {\bibinfo  {journal} {Phys. Rept.}\ }\textbf {\bibinfo {volume} {100}},\
  \bibinfo {pages} {1} (\bibinfo {year} {1983})}\BibitemShut {NoStop}%
\bibitem [{\citenamefont {Balitsky}(1996)}]{Balitsky:1995ub}%
  \BibitemOpen
  \bibfield  {author} {\bibinfo {author} {\bibfnamefont {I.}~\bibnamefont
  {Balitsky}},\ }\href {https://doi.org/10.1016/0550-3213(95)00638-9}
  {\bibfield  {journal} {\bibinfo  {journal} {Nucl. Phys. B}\ }\textbf
  {\bibinfo {volume} {463}},\ \bibinfo {pages} {99} (\bibinfo {year} {1996})},\
  \Eprint {https://arxiv.org/abs/hep-ph/9509348} {arXiv:hep-ph/9509348}
  \BibitemShut {NoStop}%
\bibitem [{\citenamefont {Kovchegov}(1999)}]{Kovchegov:1999yj}%
  \BibitemOpen
  \bibfield  {author} {\bibinfo {author} {\bibfnamefont {Y.~V.}\ \bibnamefont
  {Kovchegov}},\ }\href {https://doi.org/10.1103/PhysRevD.60.034008} {\bibfield
   {journal} {\bibinfo  {journal} {Phys. Rev. D}\ }\textbf {\bibinfo {volume}
  {60}},\ \bibinfo {pages} {034008} (\bibinfo {year} {1999})},\ \Eprint
  {https://arxiv.org/abs/hep-ph/9901281} {arXiv:hep-ph/9901281} \BibitemShut
  {NoStop}%
\bibitem [{\citenamefont {Marquet}\ \emph {et~al.}(2005)\citenamefont
  {Marquet}, \citenamefont {Peschanski},\ and\ \citenamefont
  {Soyez}}]{Marquet:2005qu}%
  \BibitemOpen
  \bibfield  {author} {\bibinfo {author} {\bibfnamefont {C.}~\bibnamefont
  {Marquet}}, \bibinfo {author} {\bibfnamefont {R.~B.}\ \bibnamefont
  {Peschanski}},\ and\ \bibinfo {author} {\bibfnamefont {G.}~\bibnamefont
  {Soyez}},\ }\href {https://doi.org/10.1016/j.nuclphysa.2005.03.089}
  {\bibfield  {journal} {\bibinfo  {journal} {Nucl. Phys. A}\ }\textbf
  {\bibinfo {volume} {756}},\ \bibinfo {pages} {399} (\bibinfo {year}
  {2005})},\ \Eprint {https://arxiv.org/abs/hep-ph/0502020}
  {arXiv:hep-ph/0502020} \BibitemShut {NoStop}%
\bibitem [{\citenamefont {Jalilian-Marian}\ \emph {et~al.}(1997)\citenamefont
  {Jalilian-Marian}, \citenamefont {Kovner}, \citenamefont {Leonidov},\ and\
  \citenamefont {Weigert}}]{JalilianMarian:1997jx}%
  \BibitemOpen
  \bibfield  {author} {\bibinfo {author} {\bibfnamefont {J.}~\bibnamefont
  {Jalilian-Marian}}, \bibinfo {author} {\bibfnamefont {A.}~\bibnamefont
  {Kovner}}, \bibinfo {author} {\bibfnamefont {A.}~\bibnamefont {Leonidov}},\
  and\ \bibinfo {author} {\bibfnamefont {H.}~\bibnamefont {Weigert}},\ }\href
  {https://doi.org/10.1016/S0550-3213(97)00440-9} {\bibfield  {journal}
  {\bibinfo  {journal} {Nucl. Phys.}\ }\textbf {\bibinfo {volume} {B504}},\
  \bibinfo {pages} {415} (\bibinfo {year} {1997})},\ \Eprint
  {https://arxiv.org/abs/hep-ph/9701284} {arXiv:hep-ph/9701284 [hep-ph]}
  \BibitemShut {NoStop}%
\bibitem [{\citenamefont {Jalilian-Marian}\ \emph
  {et~al.}(1998{\natexlab{a}})\citenamefont {Jalilian-Marian}, \citenamefont
  {Kovner}, \citenamefont {Leonidov},\ and\ \citenamefont
  {Weigert}}]{JalilianMarian:1997gr}%
  \BibitemOpen
  \bibfield  {author} {\bibinfo {author} {\bibfnamefont {J.}~\bibnamefont
  {Jalilian-Marian}}, \bibinfo {author} {\bibfnamefont {A.}~\bibnamefont
  {Kovner}}, \bibinfo {author} {\bibfnamefont {A.}~\bibnamefont {Leonidov}},\
  and\ \bibinfo {author} {\bibfnamefont {H.}~\bibnamefont {Weigert}},\ }\href
  {https://doi.org/10.1103/PhysRevD.59.014014} {\bibfield  {journal} {\bibinfo
  {journal} {Phys. Rev.}\ }\textbf {\bibinfo {volume} {D59}},\ \bibinfo {pages}
  {014014} (\bibinfo {year} {1998}{\natexlab{a}})},\ \Eprint
  {https://arxiv.org/abs/hep-ph/9706377} {arXiv:hep-ph/9706377 [hep-ph]}
  \BibitemShut {NoStop}%
\bibitem [{\citenamefont {Jalilian-Marian}\ \emph
  {et~al.}(1998{\natexlab{b}})\citenamefont {Jalilian-Marian}, \citenamefont
  {Kovner},\ and\ \citenamefont {Weigert}}]{JalilianMarian:1997dw}%
  \BibitemOpen
  \bibfield  {author} {\bibinfo {author} {\bibfnamefont {J.}~\bibnamefont
  {Jalilian-Marian}}, \bibinfo {author} {\bibfnamefont {A.}~\bibnamefont
  {Kovner}},\ and\ \bibinfo {author} {\bibfnamefont {H.}~\bibnamefont
  {Weigert}},\ }\href {https://doi.org/10.1103/PhysRevD.59.014015} {\bibfield
  {journal} {\bibinfo  {journal} {Phys. Rev.}\ }\textbf {\bibinfo {volume}
  {D59}},\ \bibinfo {pages} {014015} (\bibinfo {year} {1998}{\natexlab{b}})},\
  \Eprint {https://arxiv.org/abs/hep-ph/9709432} {arXiv:hep-ph/9709432
  [hep-ph]} \BibitemShut {NoStop}%
\bibitem [{\citenamefont {Kovner}\ \emph {et~al.}(2000)\citenamefont {Kovner},
  \citenamefont {Milhano},\ and\ \citenamefont {Weigert}}]{Kovner:2000pt}%
  \BibitemOpen
  \bibfield  {author} {\bibinfo {author} {\bibfnamefont {A.}~\bibnamefont
  {Kovner}}, \bibinfo {author} {\bibfnamefont {J.~G.}\ \bibnamefont
  {Milhano}},\ and\ \bibinfo {author} {\bibfnamefont {H.}~\bibnamefont
  {Weigert}},\ }\href {https://doi.org/10.1103/PhysRevD.62.114005} {\bibfield
  {journal} {\bibinfo  {journal} {Phys. Rev.}\ }\textbf {\bibinfo {volume}
  {D62}},\ \bibinfo {pages} {114005} (\bibinfo {year} {2000})},\ \Eprint
  {https://arxiv.org/abs/hep-ph/0004014} {arXiv:hep-ph/0004014 [hep-ph]}
  \BibitemShut {NoStop}%
\bibitem [{\citenamefont {Kovner}\ and\ \citenamefont
  {Milhano}(2000)}]{Kovner:1999bj}%
  \BibitemOpen
  \bibfield  {author} {\bibinfo {author} {\bibfnamefont {A.}~\bibnamefont
  {Kovner}}\ and\ \bibinfo {author} {\bibfnamefont {J.~G.}\ \bibnamefont
  {Milhano}},\ }\href {https://doi.org/10.1103/PhysRevD.61.014012} {\bibfield
  {journal} {\bibinfo  {journal} {Phys. Rev.}\ }\textbf {\bibinfo {volume}
  {D61}},\ \bibinfo {pages} {014012} (\bibinfo {year} {2000})},\ \Eprint
  {https://arxiv.org/abs/hep-ph/9904420} {arXiv:hep-ph/9904420 [hep-ph]}
  \BibitemShut {NoStop}%
\bibitem [{\citenamefont {Weigert}(2002)}]{Weigert:2000gi}%
  \BibitemOpen
  \bibfield  {author} {\bibinfo {author} {\bibfnamefont {H.}~\bibnamefont
  {Weigert}},\ }\href {https://doi.org/10.1016/S0375-9474(01)01668-2}
  {\bibfield  {journal} {\bibinfo  {journal} {Nucl. Phys.}\ }\textbf {\bibinfo
  {volume} {A703}},\ \bibinfo {pages} {823} (\bibinfo {year} {2002})},\ \Eprint
  {https://arxiv.org/abs/hep-ph/0004044} {arXiv:hep-ph/0004044 [hep-ph]}
  \BibitemShut {NoStop}%
\bibitem [{\citenamefont {Iancu}\ \emph {et~al.}(2001)\citenamefont {Iancu},
  \citenamefont {Leonidov},\ and\ \citenamefont {McLerran}}]{Iancu:2000hn}%
  \BibitemOpen
  \bibfield  {author} {\bibinfo {author} {\bibfnamefont {E.}~\bibnamefont
  {Iancu}}, \bibinfo {author} {\bibfnamefont {A.}~\bibnamefont {Leonidov}},\
  and\ \bibinfo {author} {\bibfnamefont {L.~D.}\ \bibnamefont {McLerran}},\
  }\href {https://doi.org/10.1016/S0375-9474(01)00642-X} {\bibfield  {journal}
  {\bibinfo  {journal} {Nucl. Phys.}\ }\textbf {\bibinfo {volume} {A692}},\
  \bibinfo {pages} {583} (\bibinfo {year} {2001})},\ \Eprint
  {https://arxiv.org/abs/hep-ph/0011241} {arXiv:hep-ph/0011241 [hep-ph]}
  \BibitemShut {NoStop}%
\bibitem [{\citenamefont {Ferreiro}\ \emph {et~al.}(2002)\citenamefont
  {Ferreiro}, \citenamefont {Iancu}, \citenamefont {Leonidov},\ and\
  \citenamefont {McLerran}}]{Ferreiro:2001qy}%
  \BibitemOpen
  \bibfield  {author} {\bibinfo {author} {\bibfnamefont {E.}~\bibnamefont
  {Ferreiro}}, \bibinfo {author} {\bibfnamefont {E.}~\bibnamefont {Iancu}},
  \bibinfo {author} {\bibfnamefont {A.}~\bibnamefont {Leonidov}},\ and\
  \bibinfo {author} {\bibfnamefont {L.}~\bibnamefont {McLerran}},\ }\href
  {https://doi.org/10.1016/S0375-9474(01)01329-X} {\bibfield  {journal}
  {\bibinfo  {journal} {Nucl. Phys. A}\ }\textbf {\bibinfo {volume} {703}},\
  \bibinfo {pages} {489} (\bibinfo {year} {2002})},\ \Eprint
  {https://arxiv.org/abs/hep-ph/0109115} {arXiv:hep-ph/0109115} \BibitemShut
  {NoStop}%
\bibitem [{\citenamefont {Navelet}\ and\ \citenamefont
  {Wallon}(1998)}]{Navelet:1997tx}%
  \BibitemOpen
  \bibfield  {author} {\bibinfo {author} {\bibfnamefont {H.}~\bibnamefont
  {Navelet}}\ and\ \bibinfo {author} {\bibfnamefont {S.}~\bibnamefont
  {Wallon}},\ }\href {https://doi.org/10.1016/S0550-3213(98)00146-1} {\bibfield
   {journal} {\bibinfo  {journal} {Nucl. Phys. B}\ }\textbf {\bibinfo {volume}
  {522}},\ \bibinfo {pages} {237} (\bibinfo {year} {1998})},\ \Eprint
  {https://arxiv.org/abs/hep-ph/9705296} {arXiv:hep-ph/9705296} \BibitemShut
  {NoStop}%
\bibitem [{\citenamefont {Ewerz}(2004)}]{Ewerz:2004}%
  \BibitemOpen
  \bibfield  {author} {\bibinfo {author} {\bibfnamefont {C.}~\bibnamefont
  {Ewerz}},\ }\href {https://doi.org/10.48550/ARXIV.HEP-PH/0403051} {\bibinfo
  {title} {The perturbative pomeron and the odderon: Where can we find them?}}
  (\bibinfo {year} {2004}),\ \Eprint {https://arxiv.org/abs/hep-ph/0403051}
  {hep-ph/0403051} \BibitemShut {NoStop}%
\bibitem [{\citenamefont {Golec-Biernat}\ and\ \citenamefont
  {Stasto}(2003)}]{Golec-Biernat:2003naj}%
  \BibitemOpen
  \bibfield  {author} {\bibinfo {author} {\bibfnamefont {K.~J.}\ \bibnamefont
  {Golec-Biernat}}\ and\ \bibinfo {author} {\bibfnamefont {A.~M.}\ \bibnamefont
  {Stasto}},\ }\href {https://doi.org/10.1016/j.nuclphysb.2003.07.011}
  {\bibfield  {journal} {\bibinfo  {journal} {Nucl. Phys. B}\ }\textbf
  {\bibinfo {volume} {668}},\ \bibinfo {pages} {345} (\bibinfo {year}
  {2003})},\ \Eprint {https://arxiv.org/abs/hep-ph/0306279}
  {arXiv:hep-ph/0306279} \BibitemShut {NoStop}%
\bibitem [{\citenamefont {Gotsman}\ and\ \citenamefont
  {Levin}(2020)}]{Levin:2020}%
  \BibitemOpen
  \bibfield  {author} {\bibinfo {author} {\bibfnamefont {E.}~\bibnamefont
  {Gotsman}}\ and\ \bibinfo {author} {\bibfnamefont {E.}~\bibnamefont
  {Levin}},\ }\href {https://doi.org/10.1103/PhysRevD.101.014023} {\bibfield
  {journal} {\bibinfo  {journal} {Phys. Rev. D}\ }\textbf {\bibinfo {volume}
  {101}},\ \bibinfo {pages} {014023} (\bibinfo {year} {2020})}\BibitemShut
  {NoStop}%
\bibitem [{\citenamefont {Gribov}(2008)}]{Gribov-diffusion}%
  \BibitemOpen
  \bibfield  {author} {\bibinfo {author} {\bibfnamefont {V.}~\bibnamefont
  {Gribov}},\ }\href {https://doi.org/10.1017/CBO9780511534942} {\emph
  {\bibinfo {title} {Strong Interactions of Hadrons at High Energies: Gribov
  Lectures on Theoretical Physics}}},\ Cambridge Monographs on Particle
  Physics, Nuclear Physics and Cosmology\ (\bibinfo  {publisher} {Cambridge
  University Press},\ \bibinfo {year} {2008})\BibitemShut {NoStop}%
\bibitem [{\citenamefont {Gribov}(1967)}]{Reggeon-Calculus:1967}%
  \BibitemOpen
  \bibfield  {author} {\bibinfo {author} {\bibfnamefont {V.~N.}\ \bibnamefont
  {Gribov}},\ }\href@noop {} {\bibfield  {journal} {\bibinfo  {journal} {Zh.
  Eksp. Teor. Fiz.}\ }\textbf {\bibinfo {volume} {53}},\ \bibinfo {pages} {654}
  (\bibinfo {year} {1967})}\BibitemShut {NoStop}%
\bibitem [{\citenamefont {Abarbanel}\ \emph {et~al.}(1975)\citenamefont
  {Abarbanel}, \citenamefont {Bronzan}, \citenamefont {Sugar},\ and\
  \citenamefont {White}}]{abarbanel1975reggeon}%
  \BibitemOpen
  \bibfield  {author} {\bibinfo {author} {\bibfnamefont {H.~D.}\ \bibnamefont
  {Abarbanel}}, \bibinfo {author} {\bibfnamefont {J.~D.}\ \bibnamefont
  {Bronzan}}, \bibinfo {author} {\bibfnamefont {R.~L.}\ \bibnamefont {Sugar}},\
  and\ \bibinfo {author} {\bibfnamefont {A.~R.}\ \bibnamefont {White}},\
  }\href@noop {} {\bibfield  {journal} {\bibinfo  {journal} {Physics Reports}\
  }\textbf {\bibinfo {volume} {21}},\ \bibinfo {pages} {119} (\bibinfo {year}
  {1975})}\BibitemShut {NoStop}%
\bibitem [{\citenamefont {Amati}\ \emph {et~al.}(1976)\citenamefont {Amati},
  \citenamefont {Le~Bellac}, \citenamefont {Marchesini},\ and\ \citenamefont
  {Ciafaloni}}]{Amati:1976}%
  \BibitemOpen
  \bibfield  {author} {\bibinfo {author} {\bibfnamefont {D.}~\bibnamefont
  {Amati}}, \bibinfo {author} {\bibfnamefont {L.}~\bibnamefont {Le~Bellac}},
  \bibinfo {author} {\bibfnamefont {G.}~\bibnamefont {Marchesini}},\ and\
  \bibinfo {author} {\bibfnamefont {M.}~\bibnamefont {Ciafaloni}},\ }\href
  {https://doi.org/https://doi.org/10.1016/0550-3213(76)90492-2} {\bibfield
  {journal} {\bibinfo  {journal} {Nuclear Physics B}\ }\textbf {\bibinfo
  {volume} {112}},\ \bibinfo {pages} {107} (\bibinfo {year}
  {1976})}\BibitemShut {NoStop}%
\bibitem [{\citenamefont {Allesandrini}\ \emph {et~al.}(1976)\citenamefont
  {Allesandrini}, \citenamefont {Amati},\ and\ \citenamefont
  {Jengo}}]{Allesandrini:1976}%
  \BibitemOpen
  \bibfield  {author} {\bibinfo {author} {\bibfnamefont {V.}~\bibnamefont
  {Allesandrini}}, \bibinfo {author} {\bibfnamefont {D.}~\bibnamefont
  {Amati}},\ and\ \bibinfo {author} {\bibfnamefont {R.}~\bibnamefont {Jengo}},\
  }\href {https://doi.org/https://doi.org/10.1016/0550-3213(76)90288-1}
  {\bibfield  {journal} {\bibinfo  {journal} {Nuclear Physics B}\ }\textbf
  {\bibinfo {volume} {108}},\ \bibinfo {pages} {425} (\bibinfo {year}
  {1976})}\BibitemShut {NoStop}%
\bibitem [{\citenamefont {Alessandrini}\ \emph {et~al.}(1977)\citenamefont
  {Alessandrini}, \citenamefont {Amati},\ and\ \citenamefont
  {Ciafaloni}}]{Alessandrini:1977}%
  \BibitemOpen
  \bibfield  {author} {\bibinfo {author} {\bibfnamefont {V.}~\bibnamefont
  {Alessandrini}}, \bibinfo {author} {\bibfnamefont {D.}~\bibnamefont
  {Amati}},\ and\ \bibinfo {author} {\bibfnamefont {M.}~\bibnamefont
  {Ciafaloni}},\ }\href {https://doi.org/10.1016/0550-3213(77)90250-4}
  {\bibfield  {journal} {\bibinfo  {journal} {Nucl. Phys. B}\ }\textbf
  {\bibinfo {volume} {130}},\ \bibinfo {pages} {429} (\bibinfo {year}
  {1977})}\BibitemShut {NoStop}%
\bibitem [{\citenamefont {Fisher}(1937)}]{FKPP:1937_A}%
  \BibitemOpen
  \bibfield  {author} {\bibinfo {author} {\bibfnamefont {R.~A.}\ \bibnamefont
  {Fisher}},\ }\href
  {https://doi.org/https://doi.org/10.1111/j.1469-1809.1937.tb02153.x}
  {\bibfield  {journal} {\bibinfo  {journal} {Annals of Eugenics}\ }\textbf
  {\bibinfo {volume} {7}},\ \bibinfo {pages} {355} (\bibinfo {year} {1937})},\
  \Eprint
  {https://arxiv.org/abs/https://onlinelibrary.wiley.com/doi/pdf/10.1111/j.1469-1809.1937.tb02153.x}
  {https://onlinelibrary.wiley.com/doi/pdf/10.1111/j.1469-1809.1937.tb02153.x}
  \BibitemShut {NoStop}%
\bibitem [{\citenamefont {Kolmogorov}\ \emph {et~al.}(1937)\citenamefont
  {Kolmogorov}, \citenamefont {Petrovsky},\ and\ \citenamefont
  {Piscounov}}]{FKPP:1937_B}%
  \BibitemOpen
  \bibfield  {author} {\bibinfo {author} {\bibfnamefont {A.}~\bibnamefont
  {Kolmogorov}}, \bibinfo {author} {\bibfnamefont {I.}~\bibnamefont
  {Petrovsky}},\ and\ \bibinfo {author} {\bibfnamefont {N.}~\bibnamefont
  {Piscounov}},\ }\href@noop {} {\bibfield  {journal} {\bibinfo  {journal}
  {Moscou Univ. Bull. Math. A 1, 1}\ }\textbf {\bibinfo {volume} {1}},\
  \bibinfo {pages} {1} (\bibinfo {year} {1937})}\BibitemShut {NoStop}%
\bibitem [{\citenamefont {Munier}\ and\ \citenamefont
  {Peschanski}(2003)}]{Munier-Peschanski:2003}%
  \BibitemOpen
  \bibfield  {author} {\bibinfo {author} {\bibfnamefont {S.}~\bibnamefont
  {Munier}}\ and\ \bibinfo {author} {\bibfnamefont {R.~B.}\ \bibnamefont
  {Peschanski}},\ }\href {https://doi.org/10.1103/PhysRevLett.91.232001}
  {\bibfield  {journal} {\bibinfo  {journal} {Phys. Rev. Lett.}\ }\textbf
  {\bibinfo {volume} {91}},\ \bibinfo {pages} {232001} (\bibinfo {year}
  {2003})},\ \Eprint {https://arxiv.org/abs/hep-ph/0309177}
  {arXiv:hep-ph/0309177} \BibitemShut {NoStop}%
\bibitem [{\citenamefont {Abazov}\ \emph {et~al.}(2021)\citenamefont {Abazov}
  \emph {et~al.}}]{Odderon:2021}%
  \BibitemOpen
  \bibfield  {author} {\bibinfo {author} {\bibfnamefont {V.~M.}\ \bibnamefont
  {Abazov}} \emph {et~al.} (\bibinfo {collaboration} {TOTEM, D0}),\ }\href
  {https://doi.org/10.1103/PhysRevLett.127.062003} {\bibfield  {journal}
  {\bibinfo  {journal} {Phys. Rev. Lett.}\ }\textbf {\bibinfo {volume} {127}},\
  \bibinfo {pages} {062003} (\bibinfo {year} {2021})},\ \Eprint
  {https://arxiv.org/abs/2012.03981} {arXiv:2012.03981 [hep-ex]} \BibitemShut
  {NoStop}%
\bibitem [{\citenamefont {Peschanski}(2009)}]{Peschanski:2009}%
  \BibitemOpen
  \bibfield  {author} {\bibinfo {author} {\bibfnamefont {R.}~\bibnamefont
  {Peschanski}},\ }\href {https://doi.org/10.1103/PhysRevD.79.105014}
  {\bibfield  {journal} {\bibinfo  {journal} {Phys. Rev. D}\ }\textbf {\bibinfo
  {volume} {79}},\ \bibinfo {pages} {105014} (\bibinfo {year}
  {2009})}\BibitemShut {NoStop}%
\bibitem [{\citenamefont {Kohara}\ \emph {et~al.}(2014)\citenamefont {Kohara},
  \citenamefont {Ferreira},\ and\ \citenamefont {Kodama}}]{kohara:2014}%
  \BibitemOpen
  \bibfield  {author} {\bibinfo {author} {\bibfnamefont {A.~K.}\ \bibnamefont
  {Kohara}}, \bibinfo {author} {\bibfnamefont {E.}~\bibnamefont {Ferreira}},\
  and\ \bibinfo {author} {\bibfnamefont {T.}~\bibnamefont {Kodama}},\ }\href
  {https://doi.org/10.1140/epjc/s10052-014-3175-x} {\bibfield  {journal}
  {\bibinfo  {journal} {Eur. Phys. J. C}\ }\textbf {\bibinfo {volume} {74}},\
  \bibinfo {pages} {3175} (\bibinfo {year} {2014})},\ \Eprint
  {https://arxiv.org/abs/1408.1599} {arXiv:1408.1599 [hep-ph]} \BibitemShut
  {NoStop}%
\bibitem [{\citenamefont {Bourrely}\ \emph {et~al.}(1979)\citenamefont
  {Bourrely}, \citenamefont {Soffer},\ and\ \citenamefont {Wu}}]{BSW:1979}%
  \BibitemOpen
  \bibfield  {author} {\bibinfo {author} {\bibfnamefont {C.}~\bibnamefont
  {Bourrely}}, \bibinfo {author} {\bibfnamefont {J.}~\bibnamefont {Soffer}},\
  and\ \bibinfo {author} {\bibfnamefont {T.~T.}\ \bibnamefont {Wu}},\ }\href
  {https://doi.org/10.1103/PhysRevD.19.3249} {\bibfield  {journal} {\bibinfo
  {journal} {Phys. Rev. D}\ }\textbf {\bibinfo {volume} {19}},\ \bibinfo
  {pages} {3249} (\bibinfo {year} {1979})}\BibitemShut {NoStop}%
\bibitem [{\citenamefont {Kovchegov}\ \emph {et~al.}(2004)\citenamefont
  {Kovchegov}, \citenamefont {Szymanowski},\ and\ \citenamefont
  {Wallon}}]{Odderon:2003}%
  \BibitemOpen
  \bibfield  {author} {\bibinfo {author} {\bibfnamefont {Y.~V.}\ \bibnamefont
  {Kovchegov}}, \bibinfo {author} {\bibfnamefont {L.}~\bibnamefont
  {Szymanowski}},\ and\ \bibinfo {author} {\bibfnamefont {S.}~\bibnamefont
  {Wallon}},\ }\href {https://doi.org/10.1016/j.physletb.2004.02.036}
  {\bibfield  {journal} {\bibinfo  {journal} {Phys. Lett. B}\ }\textbf
  {\bibinfo {volume} {586}},\ \bibinfo {pages} {267} (\bibinfo {year}
  {2004})},\ \Eprint {https://arxiv.org/abs/hep-ph/0309281}
  {arXiv:hep-ph/0309281} \BibitemShut {NoStop}%
\bibitem [{\citenamefont {Ferreira}\ \emph {et~al.}(2021)\citenamefont
  {Ferreira}, \citenamefont {Kohara},\ and\ \citenamefont
  {Kodama}}]{Kohara:2021}%
  \BibitemOpen
  \bibfield  {author} {\bibinfo {author} {\bibfnamefont {E.}~\bibnamefont
  {Ferreira}}, \bibinfo {author} {\bibfnamefont {A.~K.}\ \bibnamefont
  {Kohara}},\ and\ \bibinfo {author} {\bibfnamefont {T.}~\bibnamefont
  {Kodama}},\ }\bibfield  {journal} {\bibinfo  {journal} {The European Physical
  Journal C}\ }\textbf {\bibinfo {volume} {81}},\ \href
  {https://doi.org/10.1140/epjc/s10052-021-09056-1}
  {10.1140/epjc/s10052-021-09056-1} (\bibinfo {year} {2021})\BibitemShut
  {NoStop}%
\bibitem [{\citenamefont {Zyla}\ \emph {et~al.}(2020)\citenamefont {Zyla} \emph
  {et~al.}}]{PDG:2020}%
  \BibitemOpen
  \bibfield  {author} {\bibinfo {author} {\bibfnamefont {P.}~\bibnamefont
  {Zyla}} \emph {et~al.} (\bibinfo {collaboration} {Particle Data Group}),\
  }\href {https://doi.org/10.1093/ptep/ptaa104} {\bibfield  {journal} {\bibinfo
   {journal} {PTEP}\ }\textbf {\bibinfo {volume} {2020}},\ \bibinfo {pages}
  {083C01} (\bibinfo {year} {2020})},\ \bibinfo {note} {and 2021
  update}\BibitemShut {NoStop}%
\bibitem [{\citenamefont {Amati}\ \emph {et~al.}(1975)\citenamefont {Amati},
  \citenamefont {Caneschi},\ and\ \citenamefont {Jengo}}]{Amati:1975}%
  \BibitemOpen
  \bibfield  {author} {\bibinfo {author} {\bibfnamefont {D.}~\bibnamefont
  {Amati}}, \bibinfo {author} {\bibfnamefont {L.}~\bibnamefont {Caneschi}},\
  and\ \bibinfo {author} {\bibfnamefont {R.}~\bibnamefont {Jengo}},\ }\href
  {https://doi.org/https://doi.org/10.1016/0550-3213(75)90604-5} {\bibfield
  {journal} {\bibinfo  {journal} {Nuclear Physics B}\ }\textbf {\bibinfo
  {volume} {101}},\ \bibinfo {pages} {397} (\bibinfo {year}
  {1975})}\BibitemShut {NoStop}%
\bibitem [{\citenamefont {Selyugin}(2015)}]{Selyugin:2015}%
  \BibitemOpen
  \bibfield  {author} {\bibinfo {author} {\bibfnamefont {O.~V.}\ \bibnamefont
  {Selyugin}},\ }\href {https://doi.org/10.1103/PhysRevD.91.113003} {\bibfield
  {journal} {\bibinfo  {journal} {Phys. Rev. D}\ }\textbf {\bibinfo {volume}
  {91}},\ \bibinfo {pages} {113003} (\bibinfo {year} {2015})}\BibitemShut
  {NoStop}%
\bibitem [{\citenamefont {Csorgo}\ and\ \citenamefont
  {Szanyi}(2021)}]{Csorgo:2021}%
  \BibitemOpen
  \bibfield  {author} {\bibinfo {author} {\bibfnamefont {T.}~\bibnamefont
  {Csorgo}}\ and\ \bibinfo {author} {\bibfnamefont {I.}~\bibnamefont
  {Szanyi}},\ }\href {https://doi.org/10.1140/epjc/s10052-021-09381-5}
  {\bibfield  {journal} {\bibinfo  {journal} {Eur. Phys. J. C}\ }\textbf
  {\bibinfo {volume} {81}},\ \bibinfo {pages} {611} (\bibinfo {year} {2021})},\
  \Eprint {https://arxiv.org/abs/2005.14319} {arXiv:2005.14319 [hep-ph]}
  \BibitemShut {NoStop}%
\bibitem [{\citenamefont {Martin}(1997)}]{Martin:1997}%
  \BibitemOpen
  \bibfield  {author} {\bibinfo {author} {\bibfnamefont {A.}~\bibnamefont
  {Martin}},\ }\href@noop {} {\bibfield  {journal} {\bibinfo  {journal} {Phys.
  Lett. B}\ }\textbf {\bibinfo {volume} {404}},\ \bibinfo {pages} {137}
  (\bibinfo {year} {1997})}\BibitemShut {NoStop}%
\bibitem [{\citenamefont {Csörgő}\ and\ \citenamefont
  {Szanyi}(2022)}]{Csorgo:2022}%
  \BibitemOpen
  \bibfield  {author} {\bibinfo {author} {\bibfnamefont {T.}~\bibnamefont
  {Csörgő}}\ and\ \bibinfo {author} {\bibfnamefont {I.}~\bibnamefont
  {Szanyi}},\ }in\ \href@noop {} {\emph {\bibinfo {booktitle} {International
  Scientific Days - Femtoscopy Session}}}\ (\bibinfo {year} {2022})\ \bibinfo
  {note} {\url{https://indico.cern.ch/event/1152630/}}\BibitemShut {NoStop}%
\bibitem [{\citenamefont {Antchev}\ \emph {et~al.}(2017)\citenamefont {Antchev}
  \emph {et~al.}}]{Totem:total}%
  \BibitemOpen
  \bibfield  {author} {\bibinfo {author} {\bibfnamefont {G.}~\bibnamefont
  {Antchev}} \emph {et~al.} (\bibinfo {collaboration} {The TOTEM
  Collaboration}),\ }\href {https://doi.org/10.48550/ARXIV.1712.06153}
  {\bibinfo {title} {First measurement of elastic, inelastic and total
  cross-section at $\sqrt{s}=13$ tev by totem and overview of cross-section
  data at lhc energies}} (\bibinfo {year} {2017})\BibitemShut {NoStop}%
\bibitem [{\citenamefont {Abreu}\ \emph {et~al.}(2012)\citenamefont {Abreu}
  \emph {et~al.}}]{Auger:2012}%
  \BibitemOpen
  \bibfield  {author} {\bibinfo {author} {\bibfnamefont {P.}~\bibnamefont
  {Abreu}} \emph {et~al.} (\bibinfo {collaboration} {The Pierre Auger
  Collaboration}),\ }\href {https://doi.org/10.1103/PhysRevLett.109.062002}
  {\bibfield  {journal} {\bibinfo  {journal} {Phys. Rev. Lett.}\ }\textbf
  {\bibinfo {volume} {109}},\ \bibinfo {pages} {062002} (\bibinfo {year}
  {2012})}\BibitemShut {NoStop}%
\bibitem [{\citenamefont {et~al. (ATLAS~Collaboration)}(2014)}]{Atlas:2014}%
  \BibitemOpen
  \bibfield  {author} {\bibinfo {author} {\bibfnamefont {G.~A.}\ \bibnamefont
  {et~al. (ATLAS~Collaboration)}},\ }\href@noop {} {\bibfield  {journal}
  {\bibinfo  {journal} {Nuclear Physics B}\ }\textbf {\bibinfo {volume}
  {889}},\ \bibinfo {pages} {486} (\bibinfo {year} {2014})}\BibitemShut
  {NoStop}%
\bibitem [{\citenamefont {et~al. (ATLAS~Collaboration)}(2016)}]{Atlas:2016}%
  \BibitemOpen
  \bibfield  {author} {\bibinfo {author} {\bibfnamefont {M.~A.}\ \bibnamefont
  {et~al. (ATLAS~Collaboration)}},\ }\href@noop {} {\bibfield  {journal}
  {\bibinfo  {journal} {Physics Letters B}\ }\textbf {\bibinfo {volume}
  {761}},\ \bibinfo {pages} {158} (\bibinfo {year} {2016})}\BibitemShut
  {NoStop}%
\bibitem [{\citenamefont {Buras}\ and\ \citenamefont {Dias~de
  Deus}(1974)}]{Deus:1974}%
  \BibitemOpen
  \bibfield  {author} {\bibinfo {author} {\bibfnamefont {A.~J.}\ \bibnamefont
  {Buras}}\ and\ \bibinfo {author} {\bibfnamefont {J.}~\bibnamefont {Dias~de
  Deus}},\ }\href {https://doi.org/10.1016/0550-3213(74)90197-7} {\bibfield
  {journal} {\bibinfo  {journal} {Nucl. Phys. B}\ }\textbf {\bibinfo {volume}
  {71}},\ \bibinfo {pages} {481} (\bibinfo {year} {1974})}\BibitemShut
  {NoStop}%
\bibitem [{\citenamefont {Cs\"org\H{o}}\ \emph {et~al.}(2021)\citenamefont
  {Cs\"org\H{o}}, \citenamefont {Novak}, \citenamefont {Pasechnik},
  \citenamefont {Ster},\ and\ \citenamefont {Szanyi}}]{CsorgoScaling:2019ewn}%
  \BibitemOpen
  \bibfield  {author} {\bibinfo {author} {\bibfnamefont {T.}~\bibnamefont
  {Cs\"org\H{o}}}, \bibinfo {author} {\bibfnamefont {T.}~\bibnamefont {Novak}},
  \bibinfo {author} {\bibfnamefont {R.}~\bibnamefont {Pasechnik}}, \bibinfo
  {author} {\bibfnamefont {A.}~\bibnamefont {Ster}},\ and\ \bibinfo {author}
  {\bibfnamefont {I.}~\bibnamefont {Szanyi}},\ }\href
  {https://doi.org/10.1140/epjc/s10052-021-08867-6} {\bibfield  {journal}
  {\bibinfo  {journal} {Eur. Phys. J. C}\ }\textbf {\bibinfo {volume} {81}},\
  \bibinfo {pages} {180} (\bibinfo {year} {2021})},\ \Eprint
  {https://arxiv.org/abs/1912.11968} {arXiv:1912.11968 [hep-ph]} \BibitemShut
  {NoStop}%
\bibitem [{\citenamefont {Baldenegro}\ \emph {et~al.}(2022)\citenamefont
  {Baldenegro}, \citenamefont {Royon},\ and\ \citenamefont
  {Stasto}}]{Baldenegro:2022xrj}%
  \BibitemOpen
  \bibfield  {author} {\bibinfo {author} {\bibfnamefont {C.}~\bibnamefont
  {Baldenegro}}, \bibinfo {author} {\bibfnamefont {C.}~\bibnamefont {Royon}},\
  and\ \bibinfo {author} {\bibfnamefont {A.~M.}\ \bibnamefont {Stasto}},\
  }\href {https://doi.org/10.1016/j.physletb.2022.137141} {\bibfield  {journal}
  {\bibinfo  {journal} {Phys. Lett. B}\ }\textbf {\bibinfo {volume} {830}},\
  \bibinfo {pages} {137141} (\bibinfo {year} {2022})},\ \Eprint
  {https://arxiv.org/abs/2204.08328} {arXiv:2204.08328 [hep-ph]} \BibitemShut
  {NoStop}%
\bibitem [{\citenamefont {Dremin}(2015)}]{Dremin:2015}%
  \BibitemOpen
  \bibfield  {author} {\bibinfo {author} {\bibfnamefont {I.~M.}\ \bibnamefont
  {Dremin}},\ }\href {https://doi.org/10.3367/ufne.0185.201501d.0065}
  {\bibfield  {journal} {\bibinfo  {journal} {Physics-Uspekhi}\ }\textbf
  {\bibinfo {volume} {58}},\ \bibinfo {pages} {61} (\bibinfo {year}
  {2015})}\BibitemShut {NoStop}%
\bibitem [{\citenamefont {Arriola}\ and\ \citenamefont
  {Broniowski}(2016)}]{Broniowiski:2016}%
  \BibitemOpen
  \bibfield  {author} {\bibinfo {author} {\bibfnamefont {E.}~\bibnamefont
  {Arriola}}\ and\ \bibinfo {author} {\bibfnamefont {W.}~\bibnamefont
  {Broniowski}},\ }\href@noop {} {\bibfield  {journal} {\bibinfo  {journal}
  {Few-Body Systems}\ }\textbf {\bibinfo {volume} {57}},\ \bibinfo {pages}
  {485–490} (\bibinfo {year} {2016})},\ \Eprint
  {https://arxiv.org/abs/https://doi.org/10.48550/arXiv.1602.00288}
  {https://doi.org/10.48550/arXiv.1602.00288} \BibitemShut {NoStop}%
\bibitem [{\citenamefont {Csörg{\H{o}}}\ \emph {et~al.}(2020)\citenamefont
  {Csörg{\H{o}}}, \citenamefont {Pasechnik},\ and\ \citenamefont
  {Ster}}]{Csorgo:2020}%
  \BibitemOpen
  \bibfield  {author} {\bibinfo {author} {\bibfnamefont {T.}~\bibnamefont
  {Csörg{\H{o}}}}, \bibinfo {author} {\bibfnamefont {R.}~\bibnamefont
  {Pasechnik}},\ and\ \bibinfo {author} {\bibfnamefont {A.}~\bibnamefont
  {Ster}},\ }\bibfield  {journal} {\bibinfo  {journal} {The European Physical
  Journal C}\ }\textbf {\bibinfo {volume} {80}},\ \href
  {https://doi.org/10.1140/epjc/s10052-020-7681-8}
  {10.1140/epjc/s10052-020-7681-8} (\bibinfo {year} {2020})\BibitemShut
  {NoStop}%
\bibitem [{\citenamefont {Proch{\'{a}}zka}\ and\ \citenamefont
  {Kundr{\'{a}}t}(2020)}]{Prochzka:2020}%
  \BibitemOpen
  \bibfield  {author} {\bibinfo {author} {\bibfnamefont {J.}~\bibnamefont
  {Proch{\'{a}}zka}}\ and\ \bibinfo {author} {\bibfnamefont {V.}~\bibnamefont
  {Kundr{\'{a}}t}},\ }\bibfield  {journal} {\bibinfo  {journal} {The European
  Physical Journal C}\ }\textbf {\bibinfo {volume} {80}},\ \href
  {https://doi.org/10.1140/epjc/s10052-020-8334-7}
  {10.1140/epjc/s10052-020-8334-7} (\bibinfo {year} {2020})\BibitemShut
  {NoStop}%
\bibitem [{\citenamefont {Antchev}\ \emph {et~al.}(2020)\citenamefont {Antchev}
  \emph {et~al.}}]{Antchev_2020}%
  \BibitemOpen
  \bibfield  {author} {\bibinfo {author} {\bibfnamefont {G.}~\bibnamefont
  {Antchev}} \emph {et~al.} (\bibinfo {collaboration} {The TOTEM
  Collaboration}),\ }\bibfield  {journal} {\bibinfo  {journal} {The European
  Physical Journal C}\ }\textbf {\bibinfo {volume} {80}},\ \href
  {https://doi.org/10.1140/epjc/s10052-020-7654-y}
  {10.1140/epjc/s10052-020-7654-y} (\bibinfo {year} {2020})\BibitemShut
  {NoStop}%
\bibitem [{\citenamefont {Antchev}\ \emph {et~al.}(2012)\citenamefont {Antchev}
  \emph {et~al.}}]{TOTEM:7TeV}%
  \BibitemOpen
  \bibfield  {author} {\bibinfo {author} {\bibfnamefont {G.}~\bibnamefont
  {Antchev}} \emph {et~al.} (\bibinfo {collaboration} {TOTEM collaboration}),\
  }\href {https://doi.org/10.1209/0295-5075/101/21002} {\bibfield  {journal}
  {\bibinfo  {journal} {EPL}\ }\textbf {\bibinfo {volume} {101}},\ \bibinfo
  {pages} {21002. 12 p} (\bibinfo {year} {2012})}\BibitemShut {NoStop}%
\bibitem [{\citenamefont {Antchev}\ \emph {et~al.}(2016)\citenamefont {Antchev}
  \emph {et~al.}}]{TOTEM:8TeV}%
  \BibitemOpen
  \bibfield  {author} {\bibinfo {author} {\bibfnamefont {G.}~\bibnamefont
  {Antchev}} \emph {et~al.} (\bibinfo {collaboration} {TOTEM}),\ }\href
  {https://doi.org/10.1140/epjc/s10052-016-4399-8} {\bibfield  {journal}
  {\bibinfo  {journal} {Eur. Phys. J. C}\ }\textbf {\bibinfo {volume} {76}},\
  \bibinfo {pages} {661} (\bibinfo {year} {2016})},\ \Eprint
  {https://arxiv.org/abs/1610.00603} {arXiv:1610.00603 [nucl-ex]} \BibitemShut
  {NoStop}%
\bibitem [{\citenamefont {Antchev}\ \emph {et~al.}(2019)\citenamefont {Antchev}
  \emph {et~al.}}]{TOTEM:13TeV}%
  \BibitemOpen
  \bibfield  {author} {\bibinfo {author} {\bibfnamefont {G.}~\bibnamefont
  {Antchev}} \emph {et~al.} (\bibinfo {collaboration} {TOTEM collaboration}),\
  }\bibfield  {journal} {\bibinfo  {journal} {The European Physical Journal C}\
  }\textbf {\bibinfo {volume} {79}},\ \href
  {https://doi.org/10.1140/epjc/s10052-019-7346-7}
  {10.1140/epjc/s10052-019-7346-7} (\bibinfo {year} {2019})\BibitemShut
  {NoStop}%
\bibitem [{\citenamefont {Antchev}\ \emph {et~al.}(2022)\citenamefont {Antchev}
  \emph {et~al.}}]{Totem:8T}%
  \BibitemOpen
  \bibfield  {author} {\bibinfo {author} {\bibfnamefont {G.}~\bibnamefont
  {Antchev}} \emph {et~al.} (\bibinfo {collaboration} {TOTEM}),\ }\href
  {https://doi.org/10.1140/epjc/s10052-022-10065-x} {\bibfield  {journal}
  {\bibinfo  {journal} {Eur. Phys. J. C}\ }\textbf {\bibinfo {volume} {82}},\
  \bibinfo {pages} {263} (\bibinfo {year} {2022})},\ \Eprint
  {https://arxiv.org/abs/2111.11991} {arXiv:2111.11991 [hep-ex]} \BibitemShut
  {NoStop}%
\end{thebibliography}%

\end{document}